\documentclass[twocolumn,showpacs,preprintnumbers,aps,prl,amsmath,amssymb,floatfix,superscriptaddress]{revtex4}

\usepackage{graphicx}
\usepackage{dcolumn}
\usepackage{bm}
\usepackage{amssymb}

\begin{document}

\title{Coherent Control of Ultracold Collisions with Chirped Light: Direction Matters}

\author{M.J. Wright,$^{1,\footnote{Present address: Institut f\"ur Experimentalphysik, Universit\"at Innsbruck, Technikerstra$\ss$e 25, 6020 Innsbruck, Austria}}$ J.A. Pechkis,$^{1}$ J.L. Carini}
\affiliation{Department of Physics, University of Connecticut, Storrs, CT 06269, USA}

\author{S. Kallush,$^{2}$ R. Kosloff}
\affiliation{Department of Physical Chemistry and the Fritz Haber Research Center for Molecular Dynamics, the Hebrew University, 91094, Jerusalem, Israel}

\author{P.L. Gould}
\affiliation{Department of Physics, University of Connecticut, Storrs, CT 06269, USA}

\date{\today}

\begin{abstract}

We demonstrate the ability to coherently control ultracold atomic Rb collisions using frequency-chirped light on the nanosecond time scale. For certain center frequencies of the chirp, the rate of inelastic trap-loss collisions induced by negatively chirped light is dramatically suppressed compared to the case of a positive chirp. We attribute this to a fundamental asymmetry in the system: an excited wavepacket always moves inward on the attractive molecular potential. For a positive chirp, the resonance condition moves outward in time, while for a negative chirp, it moves inward, in the same direction as the excited wavepacket; this allows multiple interactions between the wavepacket and the light, enabling the wavepacket to be returned coherently to the ground state. Classical and quantum calculations support this interpretation.

\end{abstract}


\maketitle 

 Traditionally, the fields of ultracold physics \cite{Metcalf99} and short-pulse coherent control \cite{Rice00, Shapiro03} have followed independent paths. Besides the obvious incompatibility of sub-picosecond timescales with the motion of slow atoms, cooling typically deals with translational degrees of freedom while coherent control involves internal degrees of freedom. A potentially fruitful collaborative venture is the production of ultracold molecules by coherently-controlled photoassociation of ultracold atoms. Despite many theoretical proposals on this topic \cite{Vala00, Luc-Koenig04a, Luc-Koenig04b, Koch06a, Koch06b, Poschinger06, Koch06c, Brown06b}, experiments to date \cite{Salzmann06, Brown06a} have demonstrated coherent control of only the photo-destruction of ultracold molecules, not of their collisional formation. Here we describe our use of frequency-chirped light on the nanosecond timescale \cite{Wright05} to coherently control ultracold collisions in Rb. We observe significant differences in the collision rates for the two chirp directions. For the negative chirp, a wavepacket excited to the attractive molecular potential can subsequently be returned coherently to the ground state, thereby reducing the collision rate.

\begin{figure}
\centerline{\includegraphics[width=8.3cm]{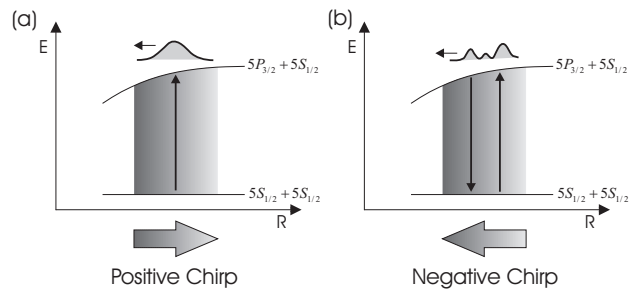}}
\caption{Schematic of ultracold collisions induced with frequency-chirped light. The ground and excited-state molecular potentials are shown at long range, as well as the excited wavepacket and the range of R (shaded region) excited by the chirp. (a) Positive (red-to-blue) chirp. The resonance condition for excitation (upward arrow) moves outward with time, while the excited wavepacket moves inward. (b) Negative (blue-to-red) chirp. The resonance condition and the wavepacket both move inward with time. Multiple interactions can return a portion of the wavepacket to the ground state (downward arrow), leading to interference.}
\end{figure}

Ultracold molecules \cite{JPhysB06} are of significant current interest due to potential applications in a variety of fields: ultracold chemistry, quantum computing, novel quantum degenerate systems, and tests of fundamental symmetries. One method of ultracold molecule production is photoassociation \cite{Jones06}, where two ultracold atoms collide in the presence of laser light and undergo a free-to-bound transition to form an excited molecule.  Spontaneous emission can subsequently produce ultracold molecules in the electronic ground state, but typically in a distribution of high vibrational levels. If the entire process could be done coherently, it might be possible to populate a single deeply-bound ground-state level. The techniques of coherent control would be useful here, since the ability to control in detail the phase and amplitude of short pulses would allow optimizing the various steps in converting pairs of free atoms into ultracold molecules in a particular quantum state. In this letter, we investigate the closely-related phenomenon of ultracold atomic collisions \cite{Weiner99}, specifically excited-state collisions resulting in trap loss. We induce these collisions with pulses of frequency-chirped light (see Fig. 1) and observe a strong dependence on the chirp direction. We attribute this difference to a fundamental asymmetry in the system \cite{Cao98}: the atom pair always moves inward on the attractive excited-state potential. Only for the negative chirp does the resonance condition also move inward, resulting in the possibility of multiple interactions between the atom pair (or collisional wavepacket) and the laser field.

The long-range interaction between ground- and excited-state atoms, separated by R, is dominated by the dipole-dipole interaction \cite{Julienne91}: -C$_{3}$/R$^{3}$. Laser light detuned from the atomic resonance by $\Delta$($\Delta$$<$0) can excite the atom pair to an attractive potential at the Condon radius R$_{c}$ = (-C$_{3}$/$\hbar\Delta$)$^{1/3}$, where the photon energy matches the potential energy difference. The initially ultracold atoms accelerate on the attractive potential and can gain sufficient energy (e.g., by predissociation from a short-range curve crossing) to escape from the trap. We investigate these trap-loss collisions by chirping over a range of frequencies, thus exciting over a range of R. We have previously shown that this chirped excitation can be adiabatic and efficient \cite{Wright05} and that there can be cooperative effects, both enhancement and depletion, between successive chirps \cite{Wright06}. In the present work, we examine the dependence of these collisions on the center detuning $\Delta$$_{c}$ of the chirp and on its direction. We observe rather different behaviors for positive and negative chirps. This is interpreted in terms of single interactions for the positive chirp versus multiple, in some cases coherent, interactions for the negative chirp. Classical and quantum calculations support this viewpoint.

We measure the trap-loss collisional rate constant $\beta$ by monitoring the density-dependent decay of a sample of ultracold ($\sim$50 $\mu$K) $^{85}$Rb atoms in a magneto-optical trap (MOT) \cite{Wright05}. Excited-state trap-loss collisions are induced by 40 ns FWHM frequency-chirped pulses with $\pm$10 GHz/$\mu$s chirp rate and 70 W/cm$^{2}$ peak intensity. This chirped light is produced with an external-cavity diode laser whose current is rapidly ramped and whose output is used to injection-lock a separate free-running diode laser \cite{Wright04}. The desired portion of each chirp is selected with an acousto-optical modulator. With the MOT light turned off for 150 $\mu$s, a number (typically 60) of these chirped pulses is applied at a repetition rate of 1.7 MHz. This entire cycle is repeated every 722 $\mu$s. The MOT repumping light is left on continuously in order to correct any possible optical pumping caused by the chirped light. Results for $\beta$ at various values of $\Delta$$_{c}$ and for both chirp directions are shown in Fig. 2. There are clearly significant differences between collisions induced with positive and negative chirps. 

\begin{figure}
\centerline{\includegraphics[width=8.3cm]{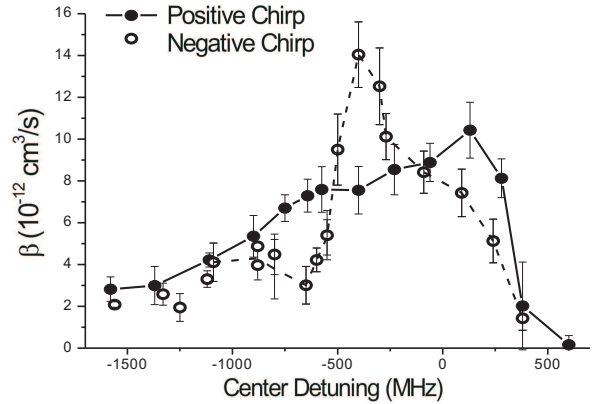}}
\caption{Collisional loss rate $\beta$ vs. $\Delta$$_{c}$/2$\pi$ for positive (solid points) and negative (open points) frequency chirps.}
\end{figure}

Focusing first on the positive chirps, we see a smooth increase in $\beta$$_{pos}$ with $\Delta$$_{c}$. This is consistent with adiabatic excitation, as illustrated in Fig. 3a. A sample atom-pair trajectory from our classical Monte-Carlo simulations (discussed below) is shown, together with R$_{c}$ as a function of time. For the positive chirp, R$_{c}$ increases with time, while the atom-pair separation R decreases. Excitation can occur when the two curves intersect. Because R and R$_{c}$ diverge after the excitation, there are no further interactions between the atom pair and the light. Returning to the dependence on center detuning (Fig. 2), as $\Delta$$_{c}$ approaches zero, the range of R$_{c}$ swept out by the chirp increases, resulting in more pairs available for excitation. The collision rate remains finite because for large R$_{c}$, the excited-state potential is very flat and it is unlikely that an atom pair will gain sufficient energy to escape before spontaneous emission occurs. For positive values of $\Delta$$_{c}$, $\beta$$_{pos}$ drops sharply because there is no longer excitation to the attractive potential.

\begin{figure}
\centerline{\includegraphics[width=8.3cm]{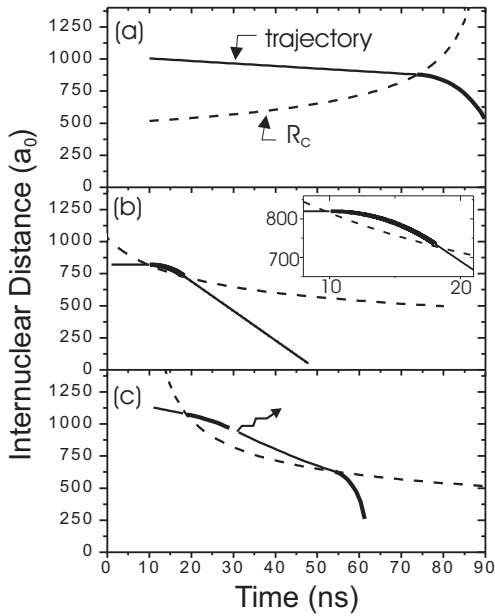}}
\caption{Sample trajectories from the classical simulations. Each plot shows the atomic separation R vs. t. The thin and thick lines indicate that the atom pair is on the ground- or excited-state potential, respectively. Also shown is the Condon radius R$_{c}$ (dashed line), which varies with time due to the chirp. The 40 ns FWHM chirped pulses are centered at t = 50 ns. (a) Adiabatic transfer for the positive chirp, $\Delta$$_{c}$/2$\pi$ = -400 MHz. (b) Coherent collision blocking for the negative chirp, $\Delta$$_{c}$/2$\pi$ = -600 MHz. The insert is an expanded view of the crossing region. The second intersection of R and R$_{c}$ gives stimulated emission back to ground state. (c) Flux enhancement for the negative chirp, $\Delta$$_{c}$/2$\pi$ = -400 MHz. The wavy arrow indicates spontaneous emission. Here, the second intersection of R and R$_{c}$ causes re-excitation.}
\end{figure}

Turning now to negative chirps, we see a more dramatic variation of $\beta$ with $\Delta$$_{c}$. In the range -1500 MHz to -1000 MHz, $\beta$$_{neg}$ is slightly below $\beta$$_{pos}$, but behaves similarly. In this regime, the excited-state potential is steep and atomic motion is fast on the time scale of the chirp. This minimizes the possibility for further interactions with the light, similar to the case of adiabatic excitation by the positive chirp.

Between -1000 MHz and -500 MHz, $\beta$$_{neg}$ is significantly suppressed relative to $\beta$$_{pos}$. Here, the time scales for the chirp and atom-pair motion on the excited potential are similar, so further interactions are possible. In particular, the excited atom pair can be stimulated back to the ground state by a second interaction with the light, effectively turning off the collision. An example of this collision blocking is shown in Fig. 3b. Since stimulated emission, as opposed to spontaneous emission, is involved, this process is coherent in nature.

In the range $\Delta$$_{c}$/2$\pi$ = -500 MHz and 0 MHz, $\beta$$_{neg}$ shows a strong peak. We attribute this to flux enhancement, as shown in Fig. 3c. Here a pair is excited at long range early in the chirp, but undergoes spontaneous emission to the flat ground-state potential before gaining significant kinetic energy. The pair is then re-excited later in the chirp, this time at short range, where the acceleration is sufficiently strong to cause trap loss. The increased collision rate is due to the attractive interaction funneling collisional flux in from long range and making it available for excitation at short range \cite{Sanchez-Villicana96}. This flux enhancement is incoherent in nature since spontaneous emission is involved. 

To model the collisions and gain insight into the physical mechanisms involved, we have performed both classical and quantum calculations. We first discuss the Monte-Carlo simulations, where the atomic motion is treated classically. Sample classical trajectories have already been shown in Fig. 3. The initial conditions (relative position vector and relative velocity vector) for an atom pair on the ground-state potential (assumed flat) are chosen randomly, in accord with a 50 $\mu$K temperature and a uniform density. This pair is then subject to the frequency-chirped pulse. When the instantaneous frequency coincides with the local resonant frequency of the pair (i.e., when R$_{c}$=R), the Landau-Zener probability for excitation to the 0$_{u}$$^{+}$ attractive molecular potential \cite{Julienne91} is calculated. If the pair is excited, it accelerates on the attractive potential until a random spontaneous emission occurs, which returns it to the ground state. If the atom pair arrives at short range (R$<$100 a$_{0}$) in the excited state, that trajectory is counted as a trap-loss event. If local resonance occurs again after the initial excitation (which can only happen for the negative chirp), the Landau-Zener transition probability is again calculated. If the atom pair is still in the excited state, this second laser-induced transition is stimulated emission back to ground state, while if the pair has decayed back to the ground state in the mean time, this second laser-induced transition is another excitation, placing the pair back on the attractive potential and giving it another chance to collide. The former case (stimulated emission) results in coherent collision blocking, while the latter case gives flux enhancement. A large number (e.g., 2x10$^{5}$) of trajectories are run and the fraction resulting in trap loss is a measure of the relative value of $\beta$.

In Fig. 4a, 
\begin{figure}
\centerline{\includegraphics[width=8.3cm]{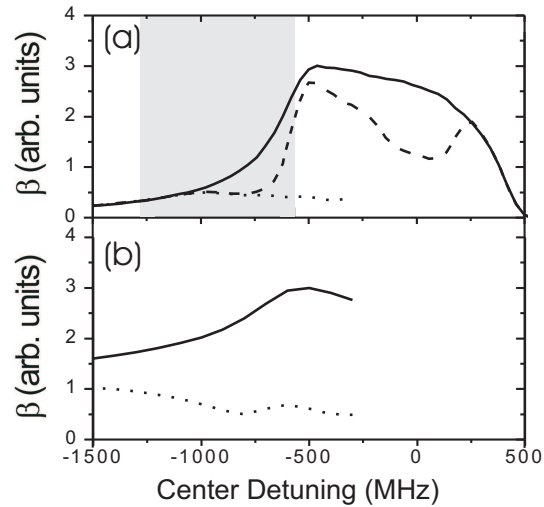}}
\caption{(a) Classical simulations of $\beta$ vs. $\Delta$$_{c}$/2$\pi$ for the positive chirp (solid line), for the negative chirp (dashed line), and for the negative chirp with flux enhancement artificially excluded (dotted line). The shaded region indicates where stimulated emission dominates over spontaneous emission for the negative chirp. (b) Quantum calculations of trap loss vs. $\Delta$$_{c}$ for positive (solid line) and negative (dotted line) chirps.}
\end{figure}
we present the dependence of $\beta$$_{pos}$ and $\beta$$_{neg}$ on $\Delta$$_{c}$. The overall trends of the data (Fig. 2) are reproduced rather well, especially considering the simplifications in the model, notably the assumption of a single excited potential and the neglect of hyperfine structure. $\beta$$_{pos}$ shows a smooth dependence on $\Delta$$_{c}$ with a rather flat region between -500 MHz and 0 MHz. $\beta$$_{neg}$ is similar to $\beta$$_{pos}$ in the range $<$-1000 MHz. For intermediate detunings, -1000 MHz to -500 MHz, the simulations show a suppression of $\beta$$_{neg}$ relative to $\beta$$_{pos}$, as seen in the data. For small negative detunings, the flux enhancement peak is reproduced. To demonstrate that the suppression is due to coherent processes, we examined individual trajectories to determine the mechanism by which the atom pair was returned to the ground state following the initial excitation. Stimulated emission indicates coherent collision blocking (Fig. 3b) while spontaneous emission indicates flux enhancement (Fig. 3c). The ratio of stimulated to spontaneous events exceeds unity in the shaded area, showing that this suppression region is clearly dominated by coherent collision blocking.

To estimate the trap loss quantum mechanically, we invoke the two-channel approximation and use the Chebychev polynomial expansion \cite{Kosloff94} with the mapped Fourier grid method \cite{Kokoouline00} to solve the time-dependent Schr\"odinger equation for $\psi$$_{g/e}$, the ground/excited electronic radial wave functions for a pair of atoms. In the rotating-wave-approximation, the dressed-state Hamiltonian is given by\begin{equation}
\setcounter{equation}{1}
\hat{H}=\hat{T}\textbf{1}+
\begin{pmatrix}
\hat{V}_g & \hbar\Omega \\ \hbar\Omega^{*} & \hat{V}_e + \Delta_c
\end{pmatrix}.
\end{equation}In Eq. (1), $\hat{T}$ is the kinetic energy, \textbf{1} is the 2x2 unit operator, $\hat{V}$$_{g/e}$ are the potential curves for the ground and the excited states, and $\Delta$$_{c}$ is the center detuning of the chirp. The light-induced coupling is $\hbar$$\Omega$=$\mu$E$_{0}$exp[-(t/${2\sigma}$)$^{2}$$\pm$i$\nu$$_{chirp}$t$^{2}$], where $\mu$ is the transition dipole moment, E$_{0}$ is the peak field amplitude, $\sigma$ is the temporal width of the pulse, and $\pm$$\nu$$_{chirp}$ is the chirp rate. The initial state $\psi$$_{g}$(t$\rightarrow$-$\infty$) is taken to be the zero energy s-wave scattering eigenstate of $\hat{V}$$_{g}$\cite{Koch06c}. Contributions to the loss from higher partial waves are found to be negligible. To model the trap loss, we add absorbing boundary conditions at the inner part (R = 100 a$_{0}$) of the excited-state potential. Any flux crossing the boundary is assumed to undergo an inelastic trap-loss collision.  To account for spontaneous decay, we couple the excited-state channel irreversibly to an artificial sink channel with characteristic lifetime $\Gamma$$^{-1}$ = 22 ns \cite{Julienne91}. Note that in this model, multiple incoherent re-excitations are forbidden, and flux enhancement does not appear. The dynamics are calculated separately for the triplet and singlet ground electronic states, adjusting each of the initial states with the proper scattering length \cite{Roberts98}. Assuming that the MOT provides unpolarized atoms, the total loss rate is a simple 3:1 statistical sum of the triplet and singlet loss rates.

The results of the fully quantum calculations for both chirp directions are shown in Fig. 4b. Since flux enhancement cannot occur in the quantum calculations, in order to more fairly compare with the classical calculations, we also show in Fig. 4a the classical result with flux enhancement trajectories artificially excluded. With this modification, the trends are in reasonable agreement. At large negative detunings, $\beta$$_{pos}$ and $\beta$$_{neg}$ converge, while at intermediate detunings, $\beta$$_{neg}$ is significantly suppressed. In the classical picture this is due to stimulated emission of the localized atom-pair, while in the quantum picture, the nonlocalized collisional wavepacket undergoes a prolonged interaction with the negatively-chirped light as it evolves. In both cases, the interaction is coherent.

In conclusion, we have demonstrated coherent control of ultracold collisions with frequency-chirped light. For certain parameters, the collision rate is dramatically suppressed simply by switching the chirp direction from positive to negative. We interpret this suppression as due to multiple coherent interactions between the collisional wavepacket and the light as the wavepacket and the resonance condition both propagate inward. Using faster time scales and larger detunings, it should be possible to form vibrational wavepackets by chirped photoassociation, an important step towards the controlled production of ground-state molecules. Incorporation of shaped pulses in lieu of simple linear chirps will allow further control and optimization of specific processes. 

The work at the University of Connecticut is supported by the Chemical Sciences, Geosciences and Biosciences Division, Office of Basic Energy Sciences, Office of Science, U.S. Department of Energy.

\end{document}